\documentclass[12pt]{article}

\usepackage{sbc-template}
\usepackage{graphicx,url}
\usepackage[utf8]{inputenc}
\usepackage[brazil]{babel}
\usepackage{amsmath,amssymb}

\usepackage{listings}

\usepackage{booktabs}

\title{QRC-Lab: An Educational Toolbox for Quantum Reservoir Computing}

\author{Anderson Fernandes Pereira dos Santos \inst{1,2}}

\address{Military Institute of Engineering (IME)\\
  Rio de Janeiro -- RJ -- Brazil
\nextinstitute
  Venturus Innovation Center\\
  Campinas -- São Paulo -- Brazil
  \email{anderson@ime.eb.br}
}

\begin{document}
\maketitle

\begin{abstract}
Quantum Reservoir Computing (QRC) has emerged as a strong paradigm for Noisy Intermediate-Scale Quantum (NISQ) machine learning, enabling the processing of temporal data with minimal training overhead by exploiting the high-dimensional dynamics of quantum states. This paper introduces \textbf{QRC-Lab}, an open-source, modular Python framework designed to bridge the gap between theoretical quantum dynamics and applied machine learning workflows. We provide a rigorous definition of QRC, contrast physical and gate-based approaches, and formalize the reservoir mapping used in the toolbox. QRC-Lab instantiates a configurable gate-based laboratory for studying input encoding, reservoir connectivity, and measurement strategies, and validates these concepts through three educational case studies: short-term memory reconstruction, temporal parity (XOR), and NARMA10 forecasting as a deliberate stress test. In addition, we include a learning-theory motivated generalization-gap scan to build intuition about capacity control in quantum feature maps. The full source code, experiment scripts, and reproducibility assets are publicly available at: \url{https://doi.org/10.5281/zenodo.18469026}.
\end{abstract}

\section{Introduction to Quantum Reservoir Computing}

The evolution of artificial intelligence and machine learning has been fundamentally driven by the increasing ability to model, analyze, and predict the behavior of dynamical systems. In particular, sequence processing and time-series analysis have historically relied on Recurrent Neural Networks (RNNs) as a dominant computational paradigm. RNNs provide a natural mathematical framework for handling temporal dependencies by maintaining an internal hidden state that recursively integrates information from past inputs. However, despite their expressive power, training RNNs has long been recognized as a challenging task. The well-known vanishing and exploding gradient problems, formally analyzed in the context of Backpropagation Through Time (BPTT), severely limit the effective learning horizon of standard RNNs \cite{hochreiter1997long}. Although gated architectures such as Long Short-Term Memory (LSTM) and Gated Recurrent Units (GRU) partially mitigate these issues, the computational and energetic cost of training deep recurrent architectures remains substantial.

A major conceptual shift occurred in the early 2000s with the independent introduction of Echo State Networks (ESNs) by Jaeger and Liquid State Machines (LSMs) by Maass. These models gave rise to the broader paradigm known as Reservoir Computing (RC) \cite{jaeger2001echo,maass2002real}. In RC, the recurrent core---the reservoir---is left untrained and randomly initialized, while only a linear readout layer is optimized. The reservoir acts as a nonlinear dynamical system that projects the input signal into a high-dimensional feature space, where temporal correlations are implicitly encoded. Under the Echo State Property, the reservoir state becomes a unique function of the recent input history, ensuring fading memory and stability. This architectural decoupling of temporal dynamics from learning leads to dramatically reduced training complexity, often replacing gradient-based optimization with simple linear regression \cite{lukosevicius2009reservoir}.

The emergence of Noisy Intermediate-Scale Quantum (NISQ) devices has naturally motivated the exploration of quantum systems as reservoirs. Quantum Reservoir Computing (QRC) extends the RC paradigm by exploiting the intrinsic dynamics of quantum systems to process temporal information \cite{fujii2017quantum}. In QRC, classical input signals are encoded into quantum states, whose evolution unfolds in a $2^N$-dimensional Hilbert space for a system of $N$ qubits. This exponential state space enables highly expressive nonlinear mappings even for relatively small quantum systems. Moreover, unlike gate-based quantum algorithms that require deep circuits and error correction, QRC is inherently compatible with the noise and decoherence present in NISQ hardware, which can be interpreted as a source of useful stochasticity rather than a limitation \cite{fujii2017quantum,nakajima2018boosting}.

Current QRC approaches can be broadly categorized into \emph{physical} and \emph{gate-based} implementations. Physical QRC leverages the natural continuous-time dynamics of specific quantum platforms, such as nuclear magnetic resonance systems, photonic reservoirs, or interacting spin ensembles. These approaches efficiently exploit native hardware dynamics but are often constrained by fixed coupling topologies and limited programmability. In contrast, gate-based QRC relies on discrete-time unitary evolutions constructed from quantum circuits, enabling fine-grained control over entanglement patterns, circuit depth, and measurement strategies. This flexibility allows systematic investigation of how architectural choices impact memory capacity, nonlinearity, and generalization performance \cite{perez2020data,chen2020temporal}.

Despite the growing body of theoretical and experimental work on QRC, a significant pedagogical gap remains. Most existing studies focus on narrow experimental setups or theoretical analyses, often accompanied by non-public or highly specialized codebases. As a result, students and researchers transitioning from foundational quantum computing concepts to quantum machine learning face a steep learning curve. In particular, there is a lack of modular, education-oriented frameworks that allow users to explore QRC principles without extensive low-level implementation overhead.

To address this gap, we introduce \emph{QRC-Lab}, a modular academic framework designed to support both education and research in gate-based Quantum Reservoir Computing. QRC-Lab provides a structured environment in which users can systematically study input encoding strategies, reservoir architectures, measurement schemes, and classical readout models. By abstracting away backend-specific details, the framework enables learners to focus on conceptual and algorithmic foundations while remaining compatible with realistic NISQ constraints.

\paragraph{Contributions.}
This paper makes the following contributions:
\begin{itemize}
    \item \textbf{(C1) Pedagogy-first toolbox:} we release \emph{QRC-Lab}, an open-source, modular gate-based framework that decomposes QRC into reusable components (encoders, reservoirs, simulator, observables, and readout), enabling controlled experimentation and classroom use.
    \item \textbf{(C2) Reproducible educational benchmarks:} we provide end-to-end scripts and artifacts for three canonical temporal tasks (short-term memory reconstruction, temporal parity/XOR, and NARMA10 forecasting), designed as a progressive teaching sequence with interpretable plots.
    \item \textbf{(C3) Capacity-control diagnostic:} we include a learning-theory motivated \emph{generalization-gap scan} (``theory scan'') that varies reservoir size and highlights the expressivity--stability trade-off in quantum feature maps.
    \item \textbf{(C4) Reproducibility package:} we publish code, experiment configurations, and figure-generation assets in a citable release (Zenodo DOI) to support auditability and re-use in education and research.
\end{itemize}

This paper provides a comprehensive presentation of \emph{QRC-Lab} as a modular toolbox for research and education in Quantum Reservoir Computing. The remainder of the paper is organized to reflect a learning-oriented workflow. Section~\ref{sec:mathematical} establishes the theoretical foundations of quantum reservoir dynamics, feature extraction, and statistical learning considerations. Section~\ref{sec:toolbox} introduces the design principles and modular software architecture of QRC-Lab, describing how these concepts are instantiated in code and how experiments can be reproduced. Section~\ref{sec:cases} presents three pedagogical case studies (memory, parity, and NARMA10 forecasting) and concludes with a risk-bound motivated generalization-gap scan intended to build intuition about capacity control. Finally, Section~\ref{sec:conclusions} discusses limitations imposed by current NISQ devices and outlines future research directions, with particular emphasis on educational workflows, pulse-level control, and heterogeneous hardware integration.

\section{The Mathematical Foundations of Quantum Reservoir Computing}
\label{sec:mathematical}

The computational relevance of Quantum Reservoir Computing (QRC) arises from its ability to transform a classical input sequence
$\{u_t\}_{t=1}^T \subset \mathbb{R}^d$ into a high-dimensional quantum dynamical trajectory
$\{\rho_t\}_{t=1}^T$ evolving in a $2^N$-dimensional Hilbert space $\mathcal{H}$.
This transformation can be formally described as a discrete-time, input-driven, dissipative quantum dynamical system, where the reservoir state is iteratively updated according to a data-dependent quantum channel \cite{fujii2017quantum,nakajima2018boosting}:
\begin{equation}
\rho_t = \mathcal{E}_{u_t}(\rho_{t-1}).
\end{equation}

Within the gate-based abstraction adopted by \emph{QRC-Lab}, this quantum channel is decomposed into two modular components: an \textbf{input encoder} $U_{\mathrm{Enc}}(u_t)$ and a fixed \textbf{reservoir evolution operator} $U_R$. The effective unitary applied at each time step is given by
\begin{equation}
U(u_t) = U_R \, U_{\mathrm{Enc}}(u_t),
\end{equation}
such that the quantum state evolves according to
\begin{equation}
|\psi_t\rangle = U(u_t) |\psi_{t-1}\rangle.
\end{equation}

The encoding unitary $U_{\mathrm{Enc}}(u_t)$ maps the classical input vector into the quantum register.
A commonly used strategy is angle encoding, implemented as a tensor product of single-qubit rotations,
\begin{equation}
U_{\mathrm{Enc}}(u_t) = \bigotimes_{i=1}^{d} R_y\!\left(f(u_{t,i})\right),
\end{equation}
where $f(\cdot)$ denotes a suitable scaling function.
To enhance the expressive power of the input map, QRC-Lab supports \emph{data re-uploading} schemes \cite{perez2020data}, in which the input is injected multiple times and interleaved with fixed random unitaries $V_\ell$, yielding a composite encoder of the form
\begin{equation}
U_{\mathrm{Enc}}(u_t) = \prod_{\ell} V_\ell \, R_y\!\left(f(u_{t,\ell})\right).
\end{equation}

The reservoir unitary $U_R$ is responsible for mixing information across qubits through entanglement and randomized dynamics.
Its design is constrained by the quantum analogue of the Echo State Property \cite{jaeger2001echo,lukosevicius2009reservoir}, which ensures that the influence of the initial state decays over time.
Consequently, the reservoir state becomes a function of a finite input window $[u_{t-W},\ldots,u_t]$, providing the fading memory required for temporal information processing.

\subsection{Feature Extraction and Observable Design}

In practice, the full quantum state $\rho_t$ cannot be accessed directly without incurring exponential overhead.
Instead, information is extracted by measuring a finite set of observables $\{O_k\}_{k=1}^{M}$, producing a classical feature vector $x_t \in \mathbb{R}^{M}$:
\begin{equation}
x_{t,k} = \mathrm{Tr}(O_k \rho_t) = \langle \psi_t | O_k | \psi_t \rangle.
\end{equation}

QRC-Lab adopts a modular observable interface.
By default, the feature set consists of local Pauli-$Z$ operators $O_k = Z_k$, yielding $M = N$ features for an $N$-qubit reservoir.
For more demanding tasks, the toolbox allows the inclusion of higher-order observables, such as two-point correlations $O_{ij} = Z_i Z_j$, which encode pairwise dependencies and partially capture the entanglement structure of the reservoir \cite{fujii2017quantum}.

The final output is computed via a classical linear readout
\begin{equation}
y_t = W_{\mathrm{out}} x_t + b,
\end{equation}
where the parameters $W_{\mathrm{out}}$ and $b$ are trained using ridge regression.
The corresponding objective function is
\begin{equation}
\mathcal{L} = \sum_t \| y_t - y_t^{\mathrm{target}} \|^2 + \alpha \| W_{\mathrm{out}} \|^2,
\end{equation}
with $\alpha$ acting as a regularization parameter.
This hybrid quantum--classical architecture preserves the exponential feature generation of the quantum reservoir while confining learning optimization to a convex problem in the classical domain, thereby avoiding gradient instability and reducing computational cost \cite{lukosevicius2009reservoir}.

\subsection{Memory Capacity and Nonlinear Expressivity}

The performance of a quantum reservoir is governed by a trade-off between \emph{linear memory capacity} and \emph{nonlinear expressivity}.
Memory capacity quantifies the extent to which past inputs $u_{t-k}$ can be reconstructed from the current feature vector $x_t$ using linear models.
In classical reservoir computing, the total memory capacity is bounded by the dimensionality of the reservoir state \cite{jaeger2001echo}.
In QRC, the exponential dimensionality of the Hilbert space suggests a potentially large memory capacity, although in practice it is limited by decoherence, measurement constraints, and the mixing properties of $U_R$ \cite{nakajima2018boosting}.

QRC-Lab provides dedicated modules for computing short-term memory (STM) metrics, enabling users to visualize how reconstruction performance decays as a function of temporal delay.
This functionality is central to the pedagogical goal of understanding how reservoir parameters control fading memory.

Nonlinear expressivity, on the other hand, characterizes the reservoir’s ability to generate rich nonlinear functions of the input history.
When combined with entangling reservoir dynamics, the measured quantum features act as an implicit nonlinear kernel \cite{perez2020data}.
This property enables small quantum systems to solve temporally nonlinear tasks, such as parity or temporal XOR problems, that are intractable for linear models of comparable size.
QRC-Lab facilitates systematic exploration of this trade-off by allowing users to vary entanglement topology, circuit depth, and observable sets within a unified experimental environment.

\subsection{Statistical Learning Theory and Risk Control}

Beyond dynamical considerations, QRC must be analyzed through the lens of statistical learning theory.
Increasing the number of qubits $N$ enlarges the hypothesis space, which improves expressive power but can also increase the risk of overfitting.
This effect can be quantified using Rademacher complexity, which measures the ability of a hypothesis class to fit random noise over a sample of size $m$ \cite{chen2020temporal, chmielewski2025riskbounds}.

For a given hypothesis $h \in \mathcal{H}$, the true risk $R(h)$ satisfies, with probability at least $1-\delta$,
\begin{equation}
R(h) \leq \hat{R}(h) + 2\mathcal{R}_m(\mathcal{H}) + \sqrt{\frac{\ln(1/\delta)}{2m}},
\end{equation}
where $\hat{R}(h)$ denotes the empirical risk.
QRC-Lab exposes this trade-off by enabling automated sweeps over reservoir size and architecture.
By jointly plotting training and test performance, users can observe how increasing Hilbert space dimensionality can reduce empirical error while simultaneously degrading generalization performance, in line with risk-bound analyses for quantum reservoirs \cite{chmielewski2025riskbounds}.

\section{The QRC-Lab Gate-Based Toolbox}
\label{sec:toolbox}

QRC-Lab is designed as a modular software toolbox rather than a monolithic quantum circuit simulator. Its architecture explicitly mirrors the academic and experimental workflow commonly adopted in the Quantum Reservoir Computing (QRC) and broader quantum machine learning literature. The guiding design principle of the toolbox is the \emph{separation of concerns}. By decomposing the QRC pipeline into independent and reusable components, QRC-Lab avoids tightly coupled implementations that hinder reproducibility and systematic analysis.

\paragraph{Open-source and reproducibility.}
To support classroom adoption and research reproducibility, QRC-Lab is distributed as an open-source repository including the full Python package, configuration files, and end-to-end scripts that reproduce all experiments and figures in this paper. The repository is available at: \url{https://doi.org/10.5281/zenodo.18469026}.

Each stage of the QRC methodology---from classical data injection to quantum evolution and classical readout---is encapsulated in a dedicated Python module. This design allows students and researchers to modify, extend, or replace individual components in isolation, enabling controlled studies of quantum--classical interfaces without rewriting the full temporal simulation logic.

\subsection{Modular Design and Software Architecture}

The QRC-Lab toolbox is organized into five primary modules, each corresponding to a conceptual stage of the QRC formalism \cite{fujii2017quantum,lukosevicius2009reservoir}. All modules are fully documented and designed to be extensible through object-oriented inheritance.

\begin{itemize}
    \item \textbf{Encoders Module}:
    Implements classical-to-quantum data injection, including angle encoding and \emph{data re-uploading} schemes \cite{perez2020data}.

    \item \textbf{Reservoirs Module}:
    Defines the fixed quantum dynamical core. The main research-oriented implementation is \lstinline{RandomCRotReservoir}, employing randomized controlled-rotation gates held fixed after initialization, promoting diverse quantum features \cite{fujii2017quantum,nakajima2018boosting}.

    \item \textbf{Simulator Module}:
    Orchestrates temporal evolution across discrete time steps. QRC-Lab supports a \lstinline{reupload\_k} mode to approximate fading memory with bounded depth under NISQ constraints \cite{chen2020temporal}.

    \item \textbf{Observables Module}:
    Defines projection into classical features and supports both ideal statevector and shots-based execution, enabling explicit study of sampling noise effects.

    \item \textbf{Readout Module}:
    Interfaces quantum features with classical learning via \lstinline{scikit-learn}, emphasizing ridge regression to stabilize learning in high-dimensional feature spaces \cite{lukosevicius2009reservoir}.
\end{itemize}

\subsection{Reproducibility, Configuration, and Figure Generation}
\label{subsec:reproducibility}

QRC-Lab is distributed with configuration files and scripts that reproduce every figure and table reported in this paper. The release archived on Zenodo includes (i) the Python package, (ii) pinned dependency metadata, (iii) YAML/JSON experiment configurations, and (iv) the generated outputs (plots and logs) for a reference run.

\paragraph{Environment.}
All experiments in this paper were executed using the QRC-Lab release associated with the Zenodo DOI above. The framework is compatible with standard scientific Python stacks (NumPy/SciPy/scikit-learn) and uses Qiskit backends for gate-based simulation. Unless otherwise stated, the reported plots were produced using an \emph{ideal} statevector backend for clarity, and a shots-based mode is also supported to illustrate sampling noise.

\paragraph{Determinism and seeds.}
For auditability, QRC-Lab exposes a global random seed that controls (i) randomized reservoir parameter initialization, (ii) data generation for synthetic benchmarks, and (iii) any randomized subsampling in training pipelines. When running in shots-based mode, randomness due to measurement sampling is controlled by the backend seed (when supported). In all cases, QRC-Lab logs the full configuration used to generate each artifact.

\paragraph{How to reproduce.}
The repository contains a single entry point for each case study and for the theory scan. In a typical installation, results can be regenerated by running, for example:
\begin{itemize}
    \item \lstinline{python scripts/run_case_memory.py --config configs/case_memory.yaml}
    \item \lstinline{python scripts/run_case_parity.py --config configs/case_parity.yaml}
    \item \lstinline{python scripts/run_case_narma10.py --config configs/case_narma10.yaml}
    \item \lstinline{python scripts/run_theory_scan.py --config configs/theory_scan.yaml}
\end{itemize}
Each script writes plots (PNG/PDF) and logs to a timestamped output folder and can optionally export the measured feature matrices for inspection.

\subsection{Hardware Transparency and Multi-Platform Roadmap}

A defining characteristic of QRC-Lab is transparency at the quantum--classical interface. Although the current stable implementation relies primarily on Qiskit for gate-based simulation, the toolbox is designed to be backend-agnostic. Core simulator and observable interfaces are decoupled from the underlying execution engine, allowing users to specify a \lstinline{backend\_config} that targets local CPU simulators, GPU-accelerated engines, or cloud-accessible quantum hardware.

\subsection{Educational Alignment and Pedagogical Objectives}

QRC-Lab was conceived as an educationally aligned toolbox to support instruction in quantum computing, machine learning, and nonlinear dynamical systems. QRC-Lab lowers barriers by replacing unstructured scripts with modular, documented components. The workflow is reflected in interactive notebooks that guide users from unitary evolution and fading memory to noisy measurement statistics and statistical learning theory, including generalization bounds \cite{chen2020temporal,chmielewski2025riskbounds}. The repository also includes ``starter'' notebooks intended for classroom settings, along with parameterized experiment templates that encourage systematic sweeps over depth, topology, observable sets, and ridge regularization.

\section{Case Studies and Educational Benchmarks}
\label{sec:cases}

To assess the versatility and robustness of the QRC-Lab toolbox, we present three representative case studies spanning increasing levels of difficulty. These experiments are organized as a \emph{pedagogical progression}: from a diagnostic that isolates fading memory, to a clean demonstration of nonlinear separability, and finally to a canonical nonlinear forecasting benchmark that highlights limitations and motivates systematic tuning.

\paragraph{Educational framing.}
Throughout Section~\ref{sec:cases}, the goal is not to maximize state-of-the-art accuracy, but to provide \emph{interpretable learning artifacts}: plots and metrics that help students connect (i) encoding depth and re-uploading, (ii) reservoir mixing and entanglement, (iii) observable design, and (iv) regularization strength, to what is observed in training and test performance. In this sense, even imperfect outcomes are valuable: controlled failures expose where a given reservoir configuration lacks memory, nonlinearity, or statistical stability.

Figures~\ref{fig:case_memory}, \ref{fig:case_narma10}, and \ref{fig:case_parity} report prediction traces for each benchmark, and Figure~\ref{fig:generalization_gap} summarizes a generalization-gap scan motivated by risk-bound intuition. Table~\ref{tab:cases} provides a compact pedagogical summary of what each case is designed to teach.

\begin{table}[t]
\centering
\caption{Pedagogical summary of the case studies. The emphasis is on what each benchmark teaches (memory, nonlinearity, and failure modes), rather than on competitive performance.}
\label{tab:cases}
\renewcommand{\arraystretch}{1.2}
\begin{tabular}{@{}p{1.1cm} p{2.6cm} p{3.2cm} p{3.4cm} p{3.2cm}@{}}
\toprule
\textbf{Case} &
\textbf{Task type} &
\textbf{Primary concept} &
\textbf{What to vary in QRC-Lab} &
\textbf{Expected outcome} \\
\midrule
1 &
Memory reconstruction &
Fading memory vs.\ mixing &
depth, topology, observables, ridge $\alpha$ &
partial tracking + clear errors \\
2 &
NARMA10 forecasting &
Long memory + nonlinearity &
re-uploading, observables ($Z$, $ZZ$), $\alpha$ &
\emph{may fail} unless tuned \\
3 &
Parity (temporal XOR) &
Nonlinear separability &
entangling dynamics, re-uploading, observables &
near-perfect decoding possible \\
\bottomrule
\end{tabular}
\end{table}

\subsection{Experimental Protocol and Default Configuration}
\label{subsec:exp_protocol}

All case studies are based on synthetic data generators included in QRC-Lab, and all experiments follow a common protocol: (i) generate a time series with a fixed seed, (ii) discard an initial washout horizon to reduce sensitivity to the initial state, (iii) build supervised pairs using a sliding-window scheme when needed, (iv) split the sequence into disjoint train/test segments, and (v) train a ridge-regression readout on the extracted quantum features. Unless explicitly stated otherwise, performance is reported using $R^2$ for regression tasks (STM and NARMA10) and accuracy for the parity classification task.

Table~\ref{tab:exp_config} summarizes the default configuration used to generate the figures in this paper. These values are intended to be \emph{educational defaults} rather than tuned optima; QRC-Lab is designed to encourage systematic sweeps over these knobs as part of classroom activities and ablation studies.

\begin{table}[t]
\centering
\caption{Default experimental configuration used to generate the case-study figures. Values are educational defaults intended for interpretability; QRC-Lab supports systematic sweeps over all parameters.}
\label{tab:exp_config}
\renewcommand{\arraystretch}{1.15}
\begin{tabular}{@{}p{3.0cm} p{9.8cm}@{}}
\toprule
\textbf{Category} & \textbf{Default setting} \\
\midrule
Backend & Ideal statevector (reference figures); optional shots-based mode (e.g., 1024 shots) \\
Randomness control & Global seed controls reservoir init + data generation; backend seed in shots mode \\
Reservoir size & $N=4$ qubits (case figures); theory scan varies $N$ \\
Reservoir depth & depth $=3$ layers of fixed random entangling blocks \\
Topology & Ring / nearest-neighbor entanglement (default); configurable \\
Encoding & Angle encoding ($R_y$); optional data re-uploading \\
Re-uploading & $k=1$ (default); increased for NARMA10 remediation studies \\
Observables & Local $Z$ (default); optional $ZZ$ correlations for feature enrichment \\
Readout & Ridge regression; $\alpha$ swept across a log grid; default $\alpha=10^{-2}$ \\
Protocol & Washout + train/test split on contiguous segments; sliding windows where applicable \\
Metrics & $R^2$ (STM, NARMA10); accuracy (parity) \\
\bottomrule
\end{tabular}
\end{table}

\subsection{Case 1: Short-Term Memory / Memory Reconstruction}
\label{subsec:case_memory}

Short-term memory (STM) benchmarks are widely adopted diagnostic tools in reservoir computing because they probe a core requirement for temporal learning: the ability to retain a compressive representation of recent inputs while still allowing the system to mix and transform information \cite{jaeger2001echo,lukosevicius2009reservoir}. In QRC-Lab, the STM task is instantiated as a memory reconstruction problem in which the target at time $t$ depends on delayed versions of the input, and the readout is trained to recover this dependence from the reservoir-generated feature vector.

\paragraph{Educational takeaway.}
This case is designed to make the \emph{fading memory} concept visible in a single plot: students should observe where predictions follow the target and where they miss rapid variations. By sweeping reservoir depth and connectivity, learners can directly see the memory--mixing trade-off: stronger mixing tends to increase feature diversity but can wash out delayed information, while weaker mixing preserves memory at the cost of expressivity.

\begin{figure}[t]
\centering
\includegraphics[width=\linewidth]{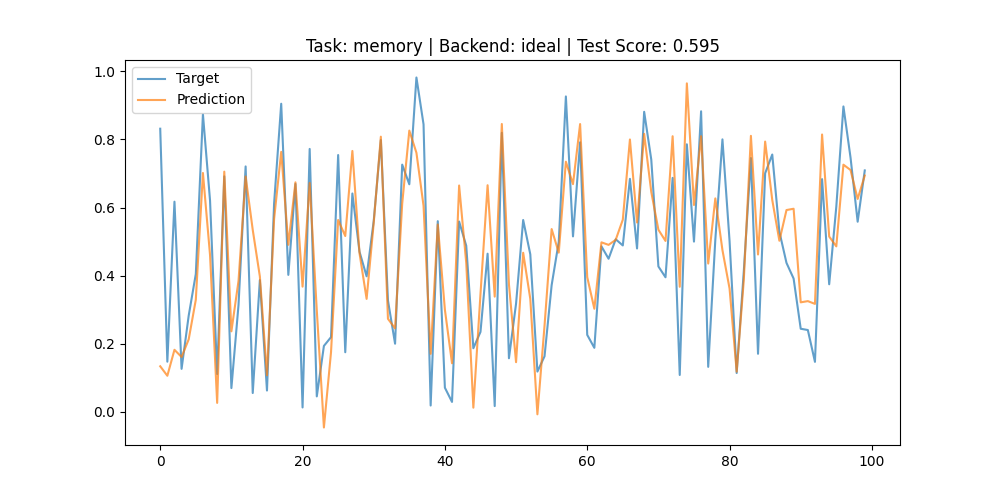}
\caption{Short-term memory / memory reconstruction task on an ideal backend. The prediction follows the target over part of the test horizon, illustrating fading memory and highlighting where dynamical mixing limits reconstruction of sharp variations.}
\label{fig:case_memory}
\end{figure}

Figure~\ref{fig:case_memory} shows a representative run for the memory task on an ideal backend. The plot overlays the target and the prediction over the test horizon, indicating that the model captures a substantial fraction of the short-term temporal structure. Sharp changes in the target are only partially tracked, which is consistent with the expected trade-off between memory retention and dynamical mixing in reservoir systems. QRC-Lab is designed to let students make this trade-off explicit by varying (i) reservoir depth, (ii) entanglement topology, (iii) the observable set, and (iv) the ridge readout regularization parameter \cite{lukosevicius2009reservoir}.

\subsection{Case 2: NARMA10 Nonlinear Forecasting (A Deliberate Stress Test)}
\label{subsec:case_narma10}

The NARMA family of benchmarks is a canonical stress test for nonlinear temporal modeling because it combines (i) long-range dependencies, (ii) nonlinear interactions, and (iii) sensitivity to the consistency of state evolution over time \cite{lukosevicius2009reservoir}. In QRC-Lab, this task is treated as one-step-ahead forecasting: given the input stream and reservoir features up to time $t$, the readout predicts the next value of the target sequence. Because the readout is linear, success depends entirely on the reservoir’s ability to embed a sufficiently rich nonlinear representation into the measured observables.

\paragraph{Educational takeaway (and why suboptimal results are useful).}
Unlike Case~1 and Case~3, NARMA10 is intentionally included as a \emph{benchmark where a naive configuration can fail}. A low or moderate test score should not be interpreted as a limitation of the toolbox; instead, it is a didactic outcome that exposes which design knobs matter for long-memory nonlinear dynamics. Students can use this case to discover that simply increasing qubits or depth does not guarantee success: the effective memory horizon, observable richness, and regularization strength must be tuned together.

\begin{figure}[t]
\centering
\includegraphics[width=\linewidth]{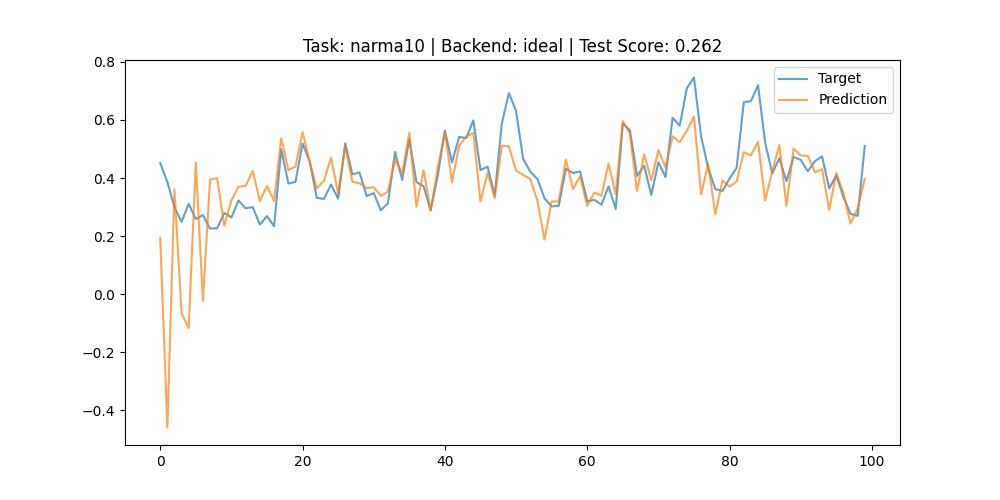}
\caption{NARMA10 nonlinear forecasting on an ideal backend. The prediction captures coarse structure but shows noticeable deviations, illustrating a common failure mode when long memory and nonlinearity are insufficiently expressed by the reservoir/observable configuration.}
\label{fig:case_narma10}
\end{figure}

Figure~\ref{fig:case_narma10} illustrates this behavior on an ideal backend. The model captures coarse trends but struggles with sharper deviations and transient behavior, a typical outcome when (i) the reservoir mixes too strongly and loses relevant delayed information or (ii) the observable set is too limited to provide independent nonlinear features to the linear readout. QRC-Lab supports systematic remedies: increasing data re-uploading depth can increase input-induced nonlinearity \cite{perez2020data}, while enriching observables (e.g., including $Z_i Z_j$ correlations) can expand the feature map without increasing qubit count. Pedagogically, this is the point: learners see the failure mode first, then reproduce the improvement through controlled ablations.

\subsection{Case 3: Parity (Temporal XOR) Classification}
\label{subsec:case_parity}

Parity (temporal XOR) is a prototypical benchmark for nonlinear sequence processing because it is not linearly separable in the raw input space and therefore requires a nonlinear transformation before a linear readout can succeed \cite{jaeger2001echo,lukosevicius2009reservoir}. In the temporal parity variant, the label at time $t$ depends on the XOR of a sliding window of recent binary inputs. This forces a reservoir to do more than remember: it must generate nonlinear combinations of past symbols that render the parity function approximately linearly separable in the feature space \cite{fujii2017quantum,nakajima2018boosting}.

\paragraph{Educational takeaway.}
This case provides a clean ``success'' example: students can see how a fixed quantum reservoir (with entangling dynamics) can generate a nonlinear feature map such that a \emph{linear} readout solves a nonlinear temporal task. In classroom use, parity is ideal for demonstrating the role of (i) entanglement patterns, (ii) re-uploading depth, and (iii) correlation observables as feature enrichers.

\begin{figure}[t]
\centering
\includegraphics[width=\linewidth]{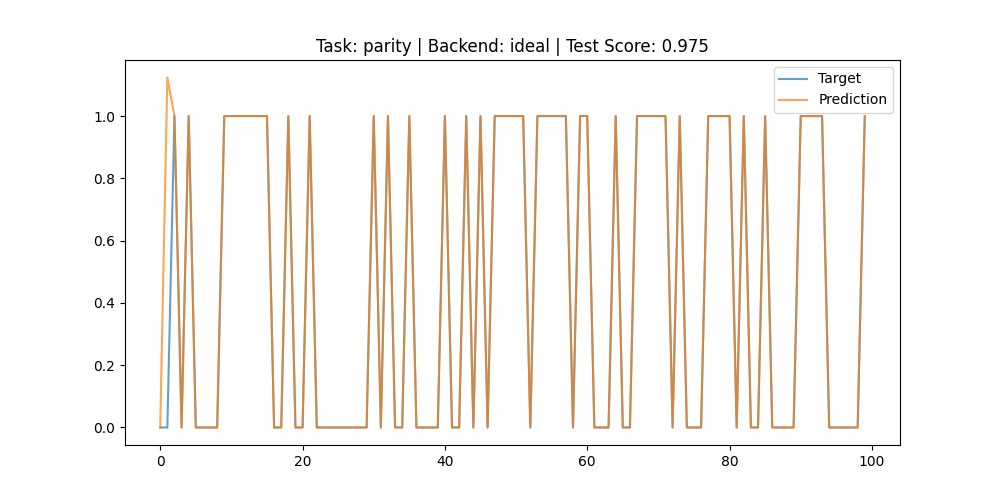}
\caption{Temporal parity (XOR) task on an ideal backend. The near-perfect overlap indicates that the quantum reservoir generates features that make a nonlinearly separable temporal rule linearly decodable by the readout.}
\label{fig:case_parity}
\end{figure}

Figure~\ref{fig:case_parity} shows an example run where the predicted sequence matches the target almost perfectly across the test horizon. QRC-Lab uses this task to teach a practical methodology: when parity is too difficult, users can systematically increase (i) the richness of the observable set (e.g., add pairwise correlations), (ii) the effective nonlinearity of injection (e.g., data re-uploading), or (iii) the reservoir mixing strength (e.g., depth or connectivity). Conversely, when parity becomes trivially perfect, QRC-Lab encourages robustness checks via noise models and finite-shot sampling, which typically degrade feature estimation and can reveal brittle solutions.

\subsection{Generalization Gap Scan and Risk-Bound Motivation}
\label{subsec:generalization_gap}

\begin{figure}[t]
\centering
\includegraphics[width=\linewidth]{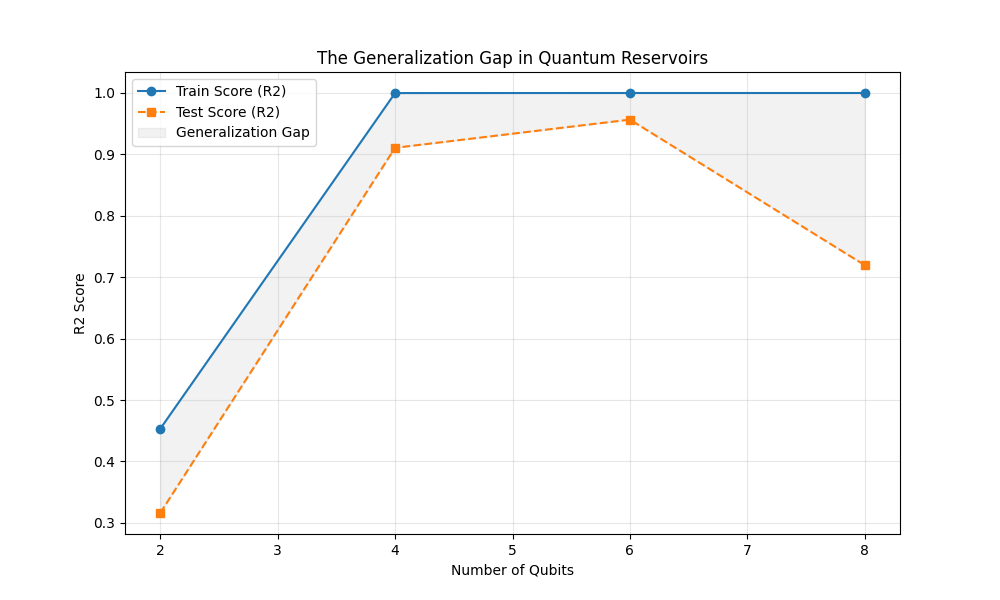}
\caption{Generalization-gap scan in QRC-Lab: training and test $R^2$ as a function of the number of qubits, highlighting diminishing returns and a regime of potential overfitting.}
\label{fig:generalization_gap}
\end{figure}

Beyond per-task performance, QRC-Lab includes a compact diagnostic called the \emph{theory scan}, which varies the number of qubits and reports training and test scores side-by-side. The resulting generalization gap curve is a practical proxy for the learning-theoretic trade-off discussed in Section~\ref{sec:mathematical}. Figure~\ref{fig:generalization_gap} illustrates this phenomenon: as the reservoir grows, the training score can saturate near its maximum, while the test score may peak and later degrade, producing a widening generalization gap. This pattern is qualitatively consistent with risk-bound analyses that relate hypothesis-class richness to Rademacher complexity in quantum reservoir families \cite{chmielewski2025riskbounds}.

\paragraph{Educational takeaway.}
This plot is primarily a conceptual visualization, not a tight statistical bound: it teaches that adding qubits can increase capacity faster than it increases generalization, unless data size and regularization are adjusted accordingly. In practice, students can replicate the scan while sweeping ridge strength and observable sets, and directly observe how regularization mitigates overfitting in high-dimensional quantum feature maps.

\section{Conclusions and Future Work}
\label{sec:conclusions}

The development and systematic evaluation of \textbf{QRC-Lab} represent a concrete step toward the consolidation of Quantum Reservoir Computing (QRC) as both an educational discipline and a viable research methodology. Throughout this work, we have presented a modular, gate-based toolbox designed to bridge the gap between abstract quantum dynamical systems and practical, data-driven temporal learning tasks. By explicitly decoupling the fundamental stages of the QRC pipeline---classical encoding, quantum reservoir evolution, measurement, and classical readout---QRC-Lab provides a transparent and reproducible environment for studying quantum-enhanced temporal processing under realistic NISQ constraints \cite{fujii2017quantum,nakajima2018boosting}.

\subsection{Summary of Contributions and Pedagogical Impact}

The primary contribution of this work is the introduction of a pedagogy-first QRC toolbox that adheres to modern software engineering principles while remaining faithful to the theoretical foundations of reservoir computing \cite{jaeger2001echo,lukosevicius2009reservoir}. Unlike monolithic and often undocumented prototypes, QRC-Lab promotes modular experimentation, enabling controlled ablation studies across encoding strategies, reservoir topologies, observable sets, and readout models.

A distinctive aspect of QRC-Lab is the explicit integration of statistical learning theory into both experimentation and pedagogy. By enabling users to visualize the generalization gap as a function of reservoir size (Figure~\ref{fig:generalization_gap}), the toolbox discourages a purely heuristic ``add more qubits'' mindset. Instead, it emphasizes the fundamental trade-off between expressivity and statistical stability, consistent with risk-bound analyses derived for quantum reservoirs \cite{chmielewski2025riskbounds}. This perspective is essential for training researchers capable of designing quantum machine learning systems that generalize reliably rather than merely fitting noise.

\subsection{Limitations and NISQ Constraints}

Despite its effectiveness as an educational and experimental platform, QRC-Lab is subject to limitations that reflect the current state of quantum computing technology. First, as a classical simulator, scalability is constrained by the exponential growth of the Hilbert space. Second, simplified noise models do not fully capture calibration drift and correlated errors. Third, practical gate-based hardware lacks native long-horizon state persistence, motivating re-uploading strategies that approximate fading memory but do not fully replicate continuous-time physical reservoirs.

\subsection{Future Directions}

Future development will focus on (i) pulse-level integration (e.g., physics-level control), (ii) broader educational benchmarking suites and notebook curricula, and (iii) hybrid/distributed reservoirs that combine multiple small quantum modules with classical communication. These directions align with the overarching goal of making QRC experimentation both scientifically grounded and educationally accessible.

\end{document}